\documentclass[showpacs,twocolumn,prb,superscriptaddress]{revtex4-1}
\usepackage{epstopdf}
\usepackage{color}
\usepackage{bm}
\usepackage{graphicx}
\usepackage{amsmath}
\usepackage{amsfonts}
\usepackage{amssymb}
\usepackage{bezier} 
\usepackage{color} 
\usepackage[colorlinks,bookmarks=false,citecolor=blue,linkcolor=red,urlcolor=blue]{hyperref}
\usepackage{times}

\makeatletter
\newsavebox{\@brx}
\newcommand{\llangle}[1][]{\savebox{\@brx}{\(\m@th{#1\langle}\)}%
  \mathopen{\copy\@brx\kern-0.5\wd\@brx\usebox{\@brx}}}
\newcommand{\rrangle}[1][]{\savebox{\@brx}{\(\m@th{#1\rangle}\)}%
  \mathclose{\copy\@brx\kern-0.5\wd\@brx\usebox{\@brx}}}
\makeatother
 
\begin{document}
\title{Numerical study of magnetization plateaux in the spin-1/2 kagome Heisenberg antiferromagnet}
\author{Sylvain Capponi}
\affiliation{Laboratoire de Physique Th\'eorique, Universit\'e de Toulouse and CNRS, UPS (IRSAMC), F-31062, Toulouse, France}
\author{Oleg Derzhko}
\affiliation{Institute for Condensed Matter Physics,
          National Academy of Sciences of Ukraine, 1 Svientsitskii Street, 
          L'viv-11, 79011, Ukraine}
\affiliation{Department for Theoretical Physics,
          Ivan Franko National University of L'viv, 12 Drahomanov Street, 
          L'viv-5, 79005, Ukraine}
\affiliation{Abdus Salam International Centre for Theoretical Physics,
          Strada Costiera 11, I-34151 Trieste, Italy}
\author{Andreas Honecker}
\affiliation{Institut f\"ur Theoretische Physik, Georg-August-Universit\"at G\"ottingen,
     D-37077 G\"ottingen, Germany}
\affiliation{Fakult\"at f\"ur Mathematik und Informatik, Georg-August-Universit\"at G\"ottingen,
     D-37073 G\"ottingen, Germany}
\author{Andreas M. L\"auchli}
\affiliation{Institut f\"ur Theoretische Physik, Universit\"at Innsbruck, A-6020 Innsbruck, Austria}
\author{Johannes Richter}
\affiliation{Institut f\"{u}r Theoretische Physik,
          Otto-von-Guericke-Universit\"{a}t Magdeburg,
          P.O. Box 4120, D-39016 Magdeburg, Germany}
\date{July 3, 2013; revised September 9, 2013}
\begin{abstract}
We clarify the existence of several magnetization plateaux for the
kagome $S=1/2$ antiferromagnetic Heisenberg model in a magnetic field.
Using approximate or exact localized magnon eigenstates, we are able
to describe in a similar manner the plateaux states that occur for
magnetization per site $m=1/3$, $5/9$, and $7/9$ of the saturation
value. These results are confirmed using large-scale Exact
Diagonalization on lattices up to 63 sites.
\end{abstract} 

\maketitle

\section{Introduction}\label{sec:introduction}

When the kagome lattice was introduced, it was shown that
the antiferromagnetic Ising model on this lattice does not order.\cite{Syozi51}
Until today, the kagome lattice remains a classic problem in highly frustrated
magnetism.\cite{HFMbook} One of the open problems concerns
the physics of the spin-1/2 Heisenberg model on the kagome
lattice in zero field, where 
despite several numerical\cite{ED,MERA,SakaiED,Lauchli2011,CCM} and variational studies,\cite{var3}
the situation is still not fully understood. The most recent DMRG studies point towards a
translationally invariant spin liquid state \cite{DMRG1,DMRG2}
with no apparent broken symmetries, a gap to triplet excitations of order $\Delta_{S=1}=0.13\,J$ and
short-range spin correlations. This state is consistent with a resonating
valence bond (RVB) state \cite{RVB} with $\mathbb{Z}_2$ topological order.\cite{DMRG2,BalentsZ2}

Returning to the Ising model on the kagome lattice, the zero-field ground state is known to be highly
degenerate.\cite{KN53} Application of a small longitudinal magnetic field polarizes one third
of the spins, but remarkably a macroscopic ground-state degeneracy survives.
Only further inclusion of quantum fluctuations lifts this degeneracy and gives
rise to a state with quantum order of valence-bond crystal (VBC) type.\cite{MoesS,one-third}
This VBC state is accompanied by a pronounced plateau in the magnetization curve
at one third of the saturation magnetization.
A similar one-third plateau was also observed in the spin-1/2 Heisenberg antiferromagnet
on the kagome lattice.\cite{Hida2001,Schulenburg2002,Richter2004,Honecker2004}
It was further argued that the states of the one-third plateau at the Heisenberg
point and close to the Ising limit belong to the same phase.\cite{one-third,Honecker2005}
Exchange anisotropy has also been shown to stabilize a 1/3 plateau in the
classical limit $S=\infty$.\cite{Cabra2002}
More recently, however, the very existence of this one-third plateau in
the spin-1/2 Heisenberg model was challenged.\cite{Nakano2010,Sakai2011a,Sakai2011b,Sakai2013}
In this context, it is noteworthy that a plateau close but not exactly equal to magnetization
$m=1/3$ has been observed experimentally in two kagome compounds.\cite{Okamoto2011}

\begin{figure}[tb!]
\centering
\includegraphics[clip=on,width=0.7\columnwidth,angle=0]{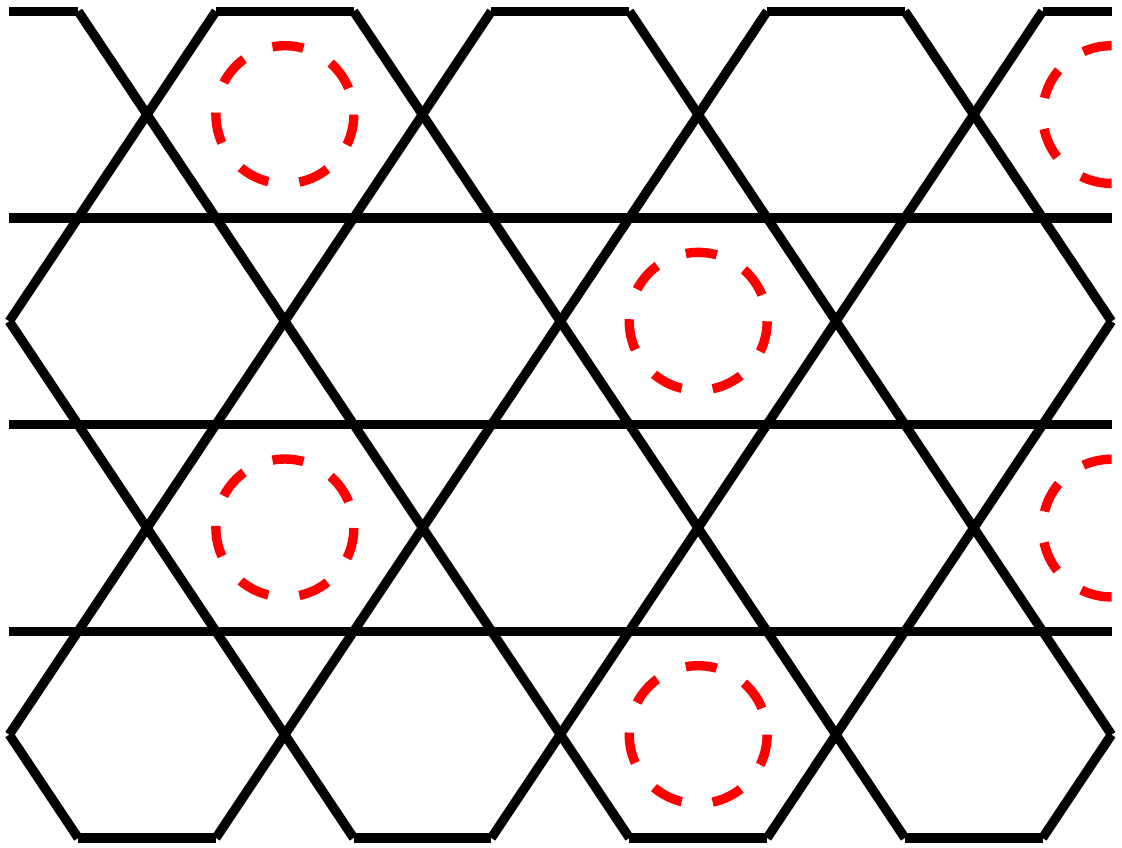}
\caption
{(Color online)
Visualization of the valence-bond-crystal type states for the $m=3/9$, $5/9$, and $7/9$ plateaux
in the $S=1/2$ Heisenberg antiferromagnet on the kagome lattice.
\label{fig01}}
\end{figure}

At very high magnetic fields, one can rigorously construct a macroscopic
number of quantum ground states for a class of highly frustrated lattices
including the kagome
lattice.\cite{Schulenburg2002,Richter2004,schmidt,ZhiTsu,loc_mag_review1,loc_mag_review2}
These exact ground states arise just below the saturation field and are
accompanied by a jump in the magnetization curve of height $\Delta m = 1/(9\,S)$
and a plateau just below this jump. For $S=1/2$, the magnetization value
on this plateaux is $m=7/9$.
Since the ground states are known exactly, one can rigorously show that
this high-field plateau exhibits the order sketched in Fig.~\ref{fig01}:
in a background of polarized (``up'') spins one flipped (``down'') spin
is localized in a quantum superposition on each hexagon marked by a dashed circle in Fig.~\ref{fig01}.
Note that the same global structure has been argued to hold at $m=1/3$,\cite{one-third,Honecker2005}
just the resonances in the hexagons are between three up and three down spins for $m=1/3$.
The states on both plateaux are indeed consistent with a two-dimensional generalization\cite{Hastings2004}
of a commensurability criterion.\cite{Oshikawa1997} This criterion would allow further
for an $m=5/9$ plateau and, indeed, the same structure as in sketched in Fig.~\ref{fig01}
suggests itself at $m=5/9$ if one now considers two down and four up spins on each hexagon.
The possibility of an $m=5/9$ plateau was mentioned previously,\cite{Honecker2004}
but has not been investigated systematically yet.

The aim of the present paper is to provide a further analysis of the 
$m=1/3$, $5/9$, and $7/9$ plateaux. 

To be concrete, we
focus on the antiferromagnetic spin-1/2 Heisenberg model on the kagome lattice
\begin{equation}\label{eq:model}
 {\cal H} = J\sum_{\langle ij\rangle} \boldsymbol{S}_i\cdot \boldsymbol{S}_j -h\sum_i S_i^z \, .
\end{equation}
$J>0$ is taken as the unit of energy and $h$ is the magnetic field along the $z$ direction. 
We will denote the magnetization per site by  $m=2\,S^z/N$,
where $S^z$ is the $z$-component of the total spin and $N$ the number of
sites. This normalization ensures a saturation value $m=1$.

Before we proceed, we mention that
similar physics arises in bosonic models on the kagome lattice,\cite{Bergman08,Huber10}
opening a further route for experimental realizations via ultracold atoms in optical
lattices.\cite{Jo12,Parameswaran2013}
%

The remainder of the manuscript is organized as follows:
In Sec.~\ref{sec:argument}, we generalize the exact wave function of the $m=7/9$ plateau to
variational VBC wave functions for the candidate plateaux at $m=1/3$ and $5/9$.
In Sec.~\ref{sec:numerics}, we perform extensive Exact Diagonalization (ED) on various lattices to confirm
(i) the existence of these three plateaux and (ii) their VBC nature.

\section{Variational model wave functions}
\label{sec:argument}

In this section we will write down variational wave functions for
the spin-1/2 model (\ref{eq:model})
on the kagome lattice at $m=1/3$ and $5/9$,
following the example of the exactly known eigenstates with $S^z \ge N/2-N/9$.%
\cite{loc_mag,Schulenburg2002,schmidt,ZhiTsu,loc_mag_review1,loc_mag_review2}
The crucial ingredients of the exact construction are ``independent''
localized magnon states where the magnons are strictly localized on the
hexagons.

The independent localized-magnon state for $S^z=7\,N/18$, i.e.,
$m=7/9$ is the so-called magnon-crystal state:
\begin{eqnarray}
\label{2.01}
\vert\Psi_{\rm VBC}^{7/9}\rangle 
= \prod_j\vert L,\downarrow \rangle_j \prod_l\vert \uparrow \rangle_l,
\end{eqnarray}
see Fig.~\ref{fig01}.
Here the first product runs over an ordered pattern of all non-overlapping
hexagons denoted by the dashed circles in Fig.~\ref{fig01}
and the second product runs over the remaining sites.
The localized-magnon state on a hexagon 
(which is the lowest-energy state of a hexagon with one spin flipped)
is 
\begin{eqnarray}
\label{2.02}
\vert L,\downarrow \rangle
= \Vert\downarrow\uparrow\uparrow\uparrow\uparrow\uparrow\rrangle_{\pi}
\, .
\end{eqnarray}
For convenience, we have introduced here the momentum eigenstate
for a hexagon
\begin{equation}
\Vert \sigma_0 \ldots \sigma_5 \rrangle_k :=
\frac{1}{\sqrt{\cal N}} \, \sum_{r=0}^{5} \exp(i\,k\,r)\,\vert \sigma_{r} \ldots
\sigma_{5+r} \rangle
\, ,
\label{eq:defKstate}
\end{equation}
where $\sigma_n = \downarrow, \uparrow$, $n+r$ has to be read modulo 6,
and ${\cal N}$ is a normalization
factor ensuring ${}_k \llangle \sigma_0 \ldots \sigma_5
\Vert \sigma_0 \ldots \sigma_5 \rrangle_k = 1$ (${\cal N} = 6$ unless
the state repeats under less than 6 translations).

One can show that
the state (\ref{2.01}), (\ref{2.02}) is not only an exact three-fold degenerate eigenstate
of the Hamiltonian ${\cal{H}}$ (\ref{eq:model}), but also a ground state
in the subspace with $S^z=7N/18$.\cite{loc_mag}
Its energy per site at $h=0$ is $e_{\rm VBC}^{7/9}/J = 1/6$.
From general arguments \cite{momoi}
the magnon-crystal state (\ref{2.01}), (\ref{2.02}) should have gapped excitations
that lead to a plateau at $m=7/9$.

On the other hand, it has been argued\cite{one-third}
that the 1/3 plateau is described by a similar wave function.
The global pattern of resonances is again as sketched in Fig.\ \ref{fig01},
but now the dashed circles represent a combination of 
the two N\'{e}el states on a hexagon. This provides a quantitative
description for the $m=1/3$ state of the $S=1/2$ XXZ Heisenberg model
in the limit of large values of the Ising anisotropy $\Delta$.\cite{one-third}
Although overlaps of the wave functions indicate
that this remains qualitatively correct for the
isotropic case $\Delta=1$, \cite{one-third}
it may still be better to consider the lowest-energy singlet state of 
the Heisenberg model on the hexagon
$\vert L,\downarrow\downarrow\downarrow \rangle$. 
Hence the corresponding three-fold degenerate valence-bond-crystal model state 
in the subspace with $S^z=3\,N/18$ reads
\begin{eqnarray}
\label{2.03}
\vert\Psi_{\rm VBC}^{3/9}\rangle
=
\prod_j\vert L,\downarrow\downarrow\downarrow \rangle_j\prod_l\vert \uparrow\rangle_l
\, .
\end{eqnarray}
The six-spin Heisenberg ring is easily diagonalized and one finds
\begin{eqnarray}
\vert L,\downarrow\downarrow\downarrow \rangle
&=& {\frac {1}{\sqrt {195+51\,\sqrt {13}}}} \Bigl( 3\,
\Vert \downarrow\downarrow\downarrow\uparrow\uparrow\uparrow\rrangle_{\pi}
\nonumber\\
&&+ {\frac {3\,(3+\sqrt {13})}{2}} \,
\left(
\Vert \downarrow\downarrow\uparrow\uparrow\downarrow\uparrow\rrangle_{\pi}
- \Vert \downarrow\downarrow\uparrow\downarrow\uparrow\uparrow\rrangle_{\pi}
\right)
\nonumber\\
&&+{\sqrt {3} \left( 4+\sqrt {13} \right) } \,
\Vert\uparrow\downarrow\uparrow\downarrow\uparrow\downarrow\rrangle_{\pi}
\Bigr) \, .
\label{2.04}
\end{eqnarray}
Although the state (\ref{2.03}), (\ref{2.04}) is not an exact eigenstate of
the Hamiltonian ${\cal{H}}$ (\ref{eq:model}), it is a good model for
the true ground state in the subspace with $S^z=3\,N/18$, see below.
The variational  energy of this state
at $h=0$ is $e_{\rm VBC}^{1/3}/J = -1/9-\sqrt {13}/18=-0.3114195153$  per site.
This is considerably lower than the {\em variational} estimate
$e_{\rm Ising}^{1/3}=0$ derived from the superposition of the two N\'eel
states.\cite{one-third} Consequently,
the state (\ref{2.03}), (\ref{2.04}) is closer to the true ground state for
$\Delta=1$.

Inspired by the valence-bond-crystal states (\ref{2.01}), (\ref{2.02})
and (\ref{2.03}), (\ref{2.04})
for the plateaux at $m=7/9$ and $m=3/9$, respectively, 
it is natural to propose a new variational wave function at $m=5/9$:
\begin{eqnarray}
\label{2.05}
\vert\Psi_{\rm VBC}^{5/9}\rangle
=
\prod_j\vert L,\downarrow\downarrow \rangle_j\prod_l\vert \uparrow \rangle_l
\end{eqnarray}
with
\begin{eqnarray}
\vert L,\downarrow\downarrow \rangle
&=&\frac{1}{2\,\sqrt{5}} \Bigl(
\left(\sqrt{5}-1\right)\,
\Vert\uparrow\uparrow\uparrow\uparrow\downarrow\downarrow\rrangle_0
\label{2.06}
\\
&&-\left(\sqrt{5}+1\right)\,
\Vert\uparrow\uparrow\uparrow\downarrow\uparrow\downarrow\rrangle_0
+2\,\sqrt{2}\,
\Vert\uparrow\uparrow\downarrow\uparrow\uparrow\downarrow\rrangle_0\Bigr) \, .
\nonumber
\end{eqnarray}
The variational energy of this state
at $h=0$ is $e_{\rm VBC}^{5/9}/J =-\sqrt {5}/18=-0.1242259987$  per site.
Again, this
is not an exact eigenstate of the Hamiltonian ${\cal{H}}$ (\ref{eq:model}),
however,
it is not far from the true ground state as extensive numerics show, see below.

It should be noted that in all three cases, the wave functions are
three-fold degenerate.
They can provide a number of consequences for correlations which
in turn can be checked by Exact Diagonalization (ED). In particular, in
view of their crystalline nature, we expect a finite gap and thus a plateau in
the magnetization curve not only for $m=7/9$, but also for $m=1/3$ and $5/9$.

In the following we present ED for finite systems 
and by comparison with theoretical predictions based on
the variational wave functions (\ref{2.03}), (\ref{2.04}) and (\ref{2.05}),
(\ref{2.06}) demonstrate that these model states
provide a good description of the physics within the plateau regimes.

\section{Numerical results}\label{sec:numerics}
We have performed extensive Exact Diagonalization (ED) using Lanczos algorithm in order to
compute the magnetization curve for various
lattices. Following Ref.~\onlinecite{Lauchli2011}, we consider a large
variety of finite lattices using periodic boundary conditions (PBC),
including also less symmetric ones that cannot accommodate the expected
VBC, in order to analyze finite-size effects. Also since the existence
of short loops going around the lattices are the major finite-size
effects, we perform the finite-size scaling using the geometric
length, i.e., the smallest distance around the torus. Definitions of
lattices and geometric distance are given in the Appendix.

\subsection{Magnetization curves} 

\begin{figure}
\includegraphics[width=\columnwidth]{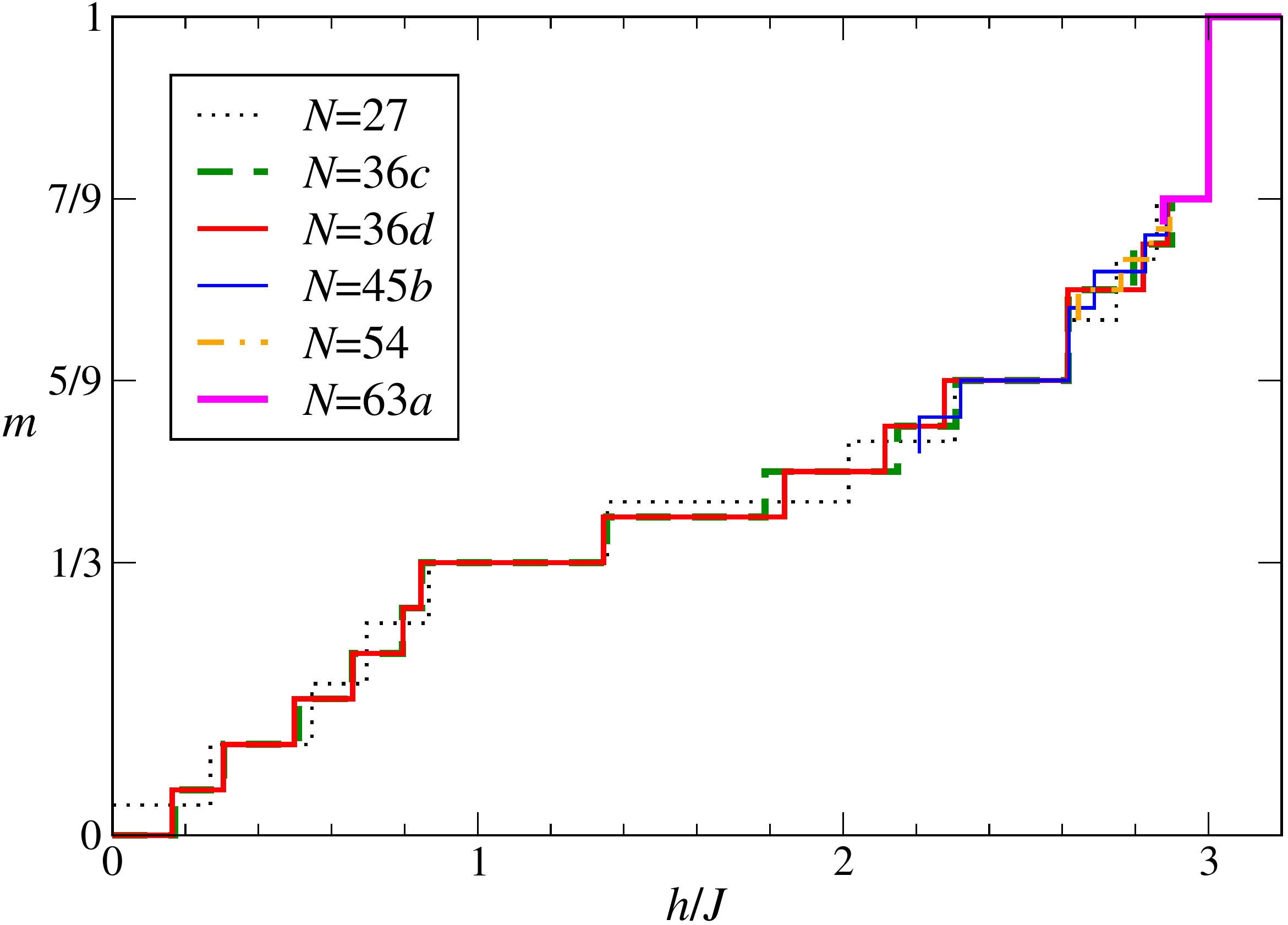}
\caption{(Color online) Magnetization curve of the $S=1/2$ kagome Heisenberg model on various lattices that can accommodate the VBC shown in Fig.~\ref{fig01} (see Appendix for details).}
\label{fig:MvsH}
\end{figure}

Since we are considering states with a large magnetization, we have to
deal with smaller Hilbert spaces than in $S^z=0$, which means that we
can access larger lattices. In this study, we have considered
lattices up to $N=63$ for which we can compute some part of the
magnetization curve. Given the number of data, we do not plot all
system sizes but Fig.~\ref{fig:MvsH} shows part of the magnetization
curves for lattices that accommodate the VBC discussed in
Sec.~\ref{sec:argument}. We recover some known features, such as the
exact saturation field $h=3J$ that can be understood in terms of the
localized magnon eigenstates and a jump to $m=7/9$. For this plateau,
we have considered more lattices than previously in the literature,
and we already see on the plot that its width seems to saturate as
system size increases. A detailed analysis will be performed
below. Similarly, looking at the $m=1/3$ and $5/9$ finite-size plateaux,
it seems that finite-size effects are rather weak both for the width
and the location of these plateaux.

\begin{figure}[tb!]
\includegraphics[width=\columnwidth]{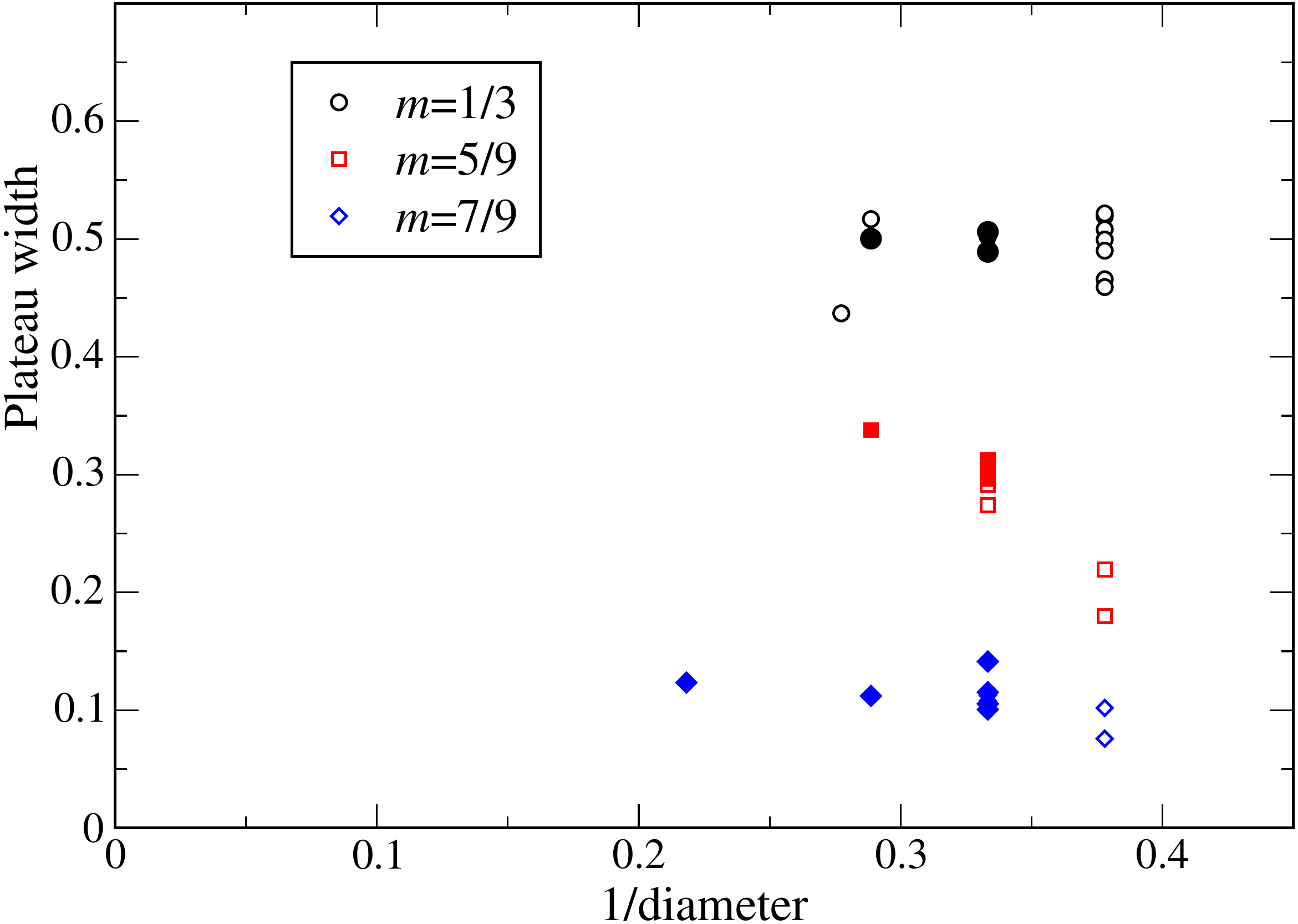}
\caption{(Color online) Widths of the $m=1/3$, $5/9$ and $7/9$
  plateaux obtained from Exact Diagonalization on various lattices (see Appendix for a list), plotted as a function of the inverse
  diameter. Filled symbols correspond to
  lattices that can accommodate the VBC states, i.e., possess the K point in their Brillouin zone: 
 $27$, $36c$, $36d$, $45b$, $54$, and $63a$). 
Open symbols are for the others. 
Data are consistent with finite  values for all three plateaux in the thermodynamic limit.}
\label{fig:width}
\end{figure}

Focusing on the expected plateaux at $m=1/3$, $5/9$, and $7/9$, we plot
their widths in Fig.~\ref{fig:width} as a function of the inverse
diameter which we believe is the relevant parameter. First of all, we
do observe some variations of the data and peculiar results for the
smallest lattices, but if we rely on the largest
lattices (in the sense of their diameter, see Appendix), 
then we do observe a tendency to saturation to finite values for
all three plateaux. Moreover, since our scenario relies on the
existence of VBC states that \emph{do not fit} on all lattices (for instance,
the unit cell  has 9 sites so clusters need to have $9p$ sites), it is
not expected a priori to have similar widths on different lattices,
which could explain some scattering in the data. Therefore, we
consider only lattices having the K point in the Brillouin zone (see inset of Fig.~\ref{fig:gaps} for a plot of the Brillouin zone).

In order to perform finite-size scaling, it does not seem appropriate
to us to use a simple linear fit extrapolation as done in
Ref.~\onlinecite{Sakai2011a}. Indeed, for a finite plateau, one expects
an exponential saturation when the system size (or diameter)
increases. Hence, by performing such extrapolation of our data, we
obtain \emph{finite} plateaux for $m=1/3$, $5/9$, and $7/9$, with decreasing
widths of the order of $0.5\, J$, $0.3\, J$ and $0.1\, J$,
respectively. For $m=7/9$, this new estimate for the plateau width
is actually slightly larger than a previous estimate\cite{spin-peierls}
$0.07\,J$.

\subsection{Energetics}

\begin{figure}[tb!]
\includegraphics[width=\columnwidth]{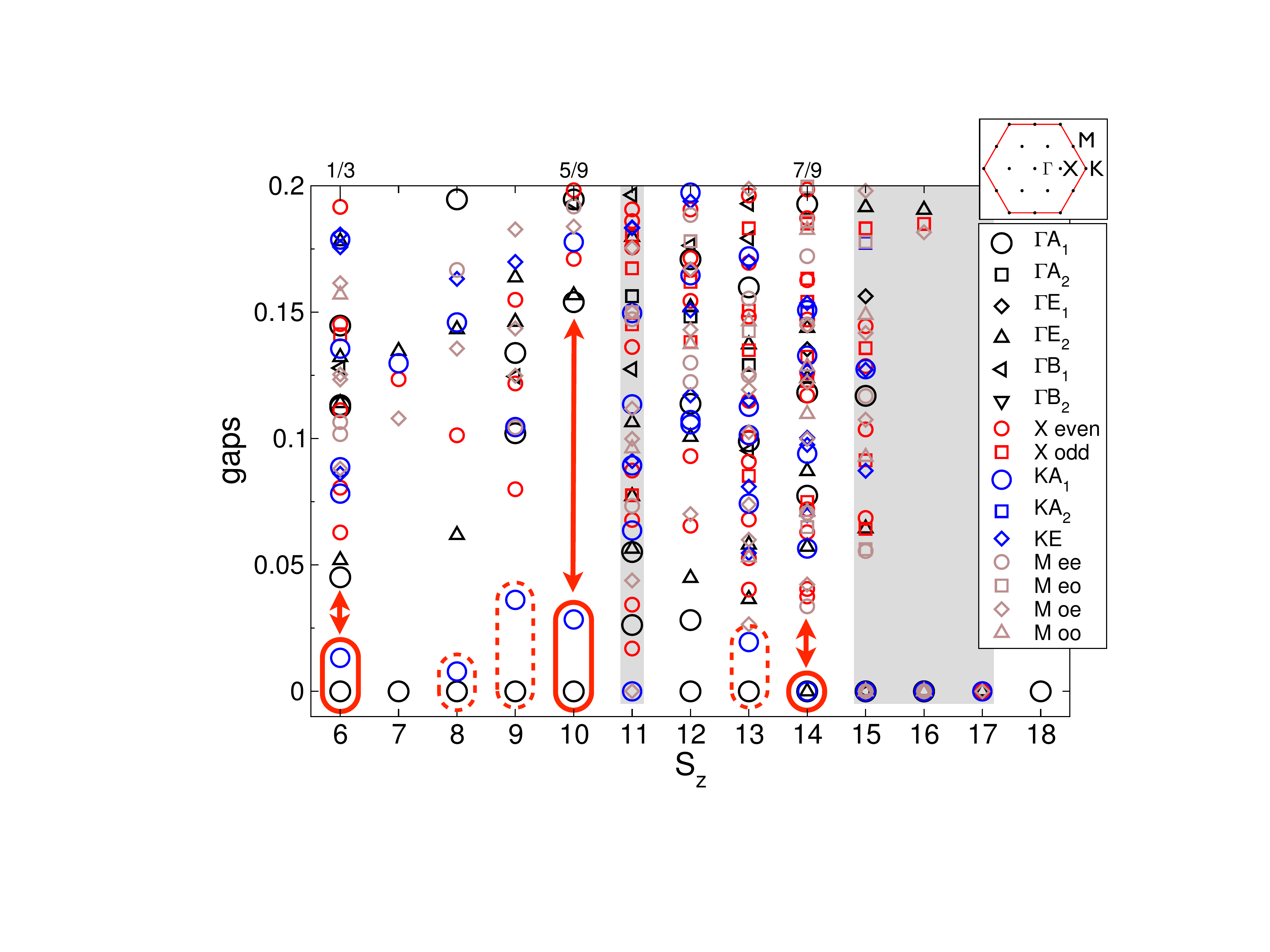}
\caption{(Color online) Energy gaps vs $S^z$ for $N=36d$ lattice labeled with
their quantum numbers. For $S^z=6$, $10$ and $14$ corresponding respectively to
$m=1/3$, $5/9$ and $7/9$, the lowest excitation corresponds to a two-fold degenerate
state at the K point in the Brillouin zone, and then a sizable gap above it (indicated by the encircled symbols
and the arrows). These spectra are compatible with the VBC states for such magnetizations. The dashed 
encircled symbols denote levels which might indicate persistence of symmetry breaking also away from the 
nominal plateaux. The magnetization sectors with a gray background are not
visited in the magnetization curve, i.e., 
they are obscured by a magnetization jump.}
\label{fig:gaps}
\end{figure}

In order to characterize possible symmetry breaking, it is useful to
investigate the low-energy levels quantum numbers on a finite
lattice. For instance, according to Sec.~\ref{sec:argument}, we expect
to have VBC states on these plateaux with three-fold degeneracy. In
the thermodynamic limit, this implies that we have degeneracy between states
at the $\Gamma$ point $(\Gamma A_1)$ and the two-fold degenerate K point (K $A_1$). In
Fig.~\ref{fig:gaps}, we plot the energy gaps obtained by computing the
ten lowest eigenstates in each symmetry sector using a Davidson
algorithm. We have subtracted the ground-state energy for each $S^z$, but for
comparison with the variational VBC states, let us mention that on $N=36d$
lattice, the ground-state energy per site for $m=1/3$ is $e_0=-0.347711\,J$, for
$m=5/9$: $e_0=-0.137251\,J$, and for $m=7/9$ we get the exact VBC state with
$e_0=J/6$. Therefore, our simple VBC wavefunctions (without any adjustable parameter) already give a reasonable estimate of these ground-state energies. 

About the excited states shown in Fig.~\ref{fig:gaps}, 
exact degenerate magnon eigenstates are found for $m\geq
7/9$ as expected. For $m=7/9$ (i.e., $S^z=14$), the ground-state is
more than three-fold degenerate due to small loops going around the
lattice (on $N=63a$, degeneracy is exactly three, see below), but there is evidence of a
small gap above them. At $m=5/9$ ($S^z=10$), we do observe two-fold
degenerate states with momentum K close to the $\Gamma$ ground-state, and
a sizable gap above them.

Let us also mention a possible feature below the $m=5/9$ plateau: on
the low-energy spectrum, we observe the same feature as on the
plateau, which could signal the persistence of VBC order \emph{away} from this magnetization. 
However,
since we expect that the magnetization decreases smoothly from the
plateau, such a state could possibly exhibit both off-diagonal and diagonal
long-range order, i.e., a supersolid state.\cite{supersolid} One has to be
cautious about this scenario, since other possibilities exist such as the
absence of superfluid signal, or magnetization jump. Nevertheless, we believe
that this would be an interesting topic to investigate further. For instance,
simple bosonic models on the same lattice only exhibit plateaux at
$1/3$, and there is no supersolid phase.\cite{Isakov06}

As a conclusion on this part, we have shown 
that the low-energy spectrum points towards a VBC scenario, that we will now confirm by directly computing relevant correlations.

\subsection{Correlations}\label{sec:correlations}

\begin{figure*}
\includegraphics[width=\linewidth]{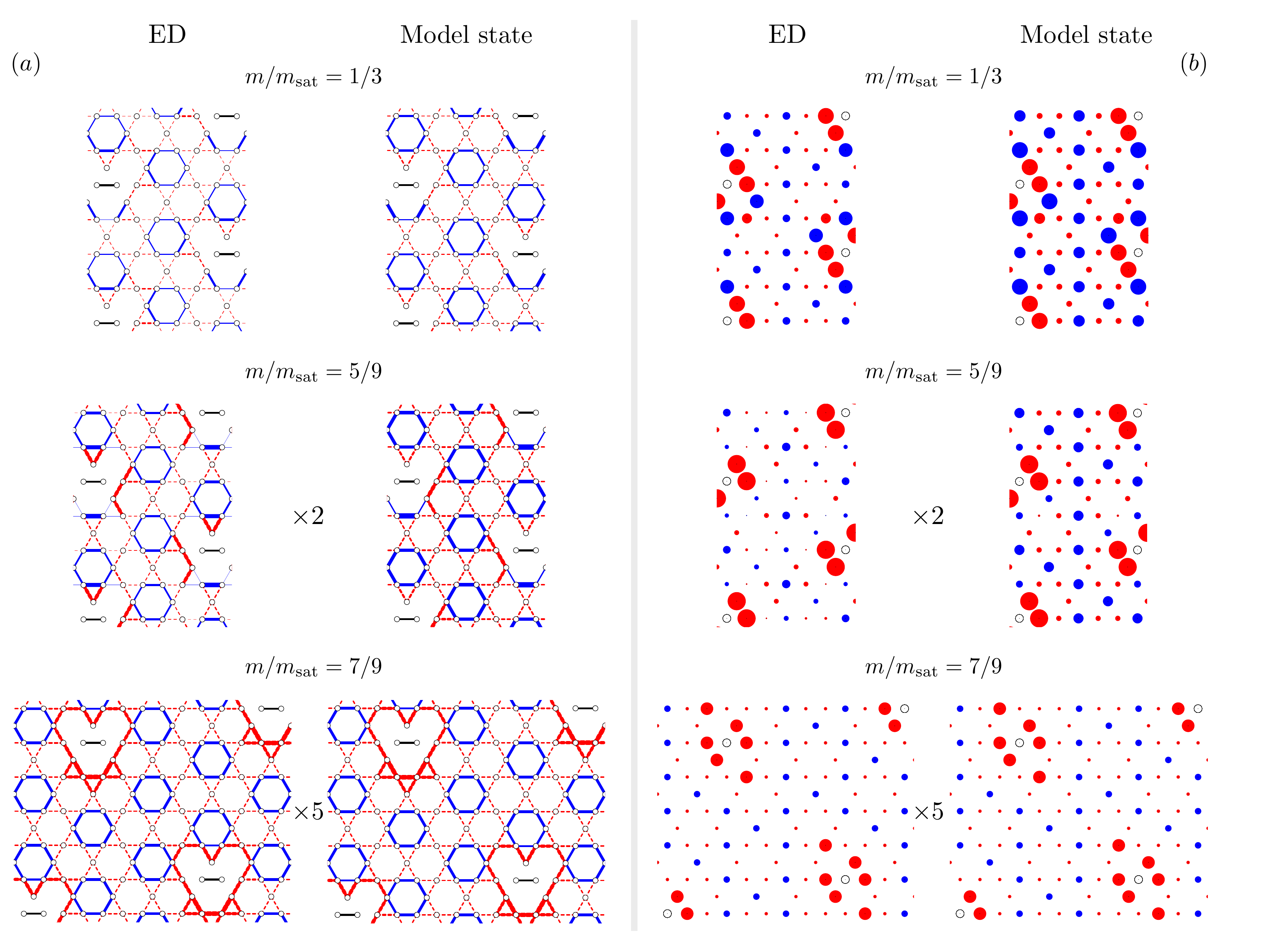}
\caption{(Color online) Dimer and spin correlations (see text for definitions) computed either by ED or on the magnon VBC state for various magnetizations: on $N=36d$ for $m=1/3$ and $5/9$, on $N=63a$ for $m=7/9$. (a) Dimer-dimer correlations (cf.~Eq.~(\ref{eq:dimerdimer})): positive and negative values are shown respectively with filled blue lines
(respectively dashed red lines) and width is proportional to the data (see Table~\ref{table2}); reference bond is shown in black. (b) Spin correlations (cf.~Eq.~(\ref{eq:szsz})): positive and negative values are shown respectively with filled blue (resp. red) disks and diameter is proportional to the data (see Table~\ref{table1}); the reference site is shown as an empty black circle. In order to have similar amplitudes, scale is multiplied by 2 and 5 for $m=5/9$ and $7/9$ with respect to $m=1/3$ data.}
\label{fig:CorrPanel}
\end{figure*}

Having established the existence of these three plateaux, we now turn
to their characterization. Let us remind that according to Hastings'
theorem,\cite{Hastings2004} $m=5/9$ and $m=7/9$ plateaux necessarily
correspond to a (at least three-fold) degenerate ground-state. 
While exotic scenarios with
topological degeneracy are possible, the more usual case is to have a
system that breaks lattice symmetries. This is not necessarily the case for
$m=1/3$, but our arguments (see Sec.~\ref{sec:argument}) indicate that
all three plateaux correspond to similar three-fold degenerate VBC states.

For each magnetization $m$, we have computed connected spin correlation functions 
\begin{equation}\label{eq:szsz}
\langle S_i^z S_j^z\rangle_c = \langle S_i^z S_j^z\rangle - \langle
S_i^z\rangle \langle S_j^z\rangle
\end{equation}
as well as connected dimer-dimer correlations
\begin{eqnarray}\label{eq:dimerdimer}
&&\langle (\boldsymbol{S}_i\cdot \boldsymbol{S}_j) (\boldsymbol{S}_k\cdot \boldsymbol{S}_\ell) \rangle_c  \nonumber\\
&&= \langle (\boldsymbol{S}_i\cdot \boldsymbol{S}_j) (\boldsymbol{S}_k\cdot \boldsymbol{S}_\ell) \rangle  - \langle (\boldsymbol{S}_i\cdot \boldsymbol{S}_j)\rangle \langle (\boldsymbol{S}_k\cdot \boldsymbol{S}_\ell) \rangle
\end{eqnarray}
using ED on the $N=36d$ lattice for $m=1/3$ and $5/9$, and the $N=63a$ lattice
for $m=7/9$. These are the largest lattices (in the sense of their
diameter, see Appendix) available for each $m$ value.  Data are
presented in Fig.~\ref{fig:CorrPanel}.

In order to make a more precise connection with our VBC picture, we
have also computed the same quantities on the pure states as shown in
Fig.~\ref{fig01}. For each $m$, we have three degenerate states that
are orthogonal in the thermodynamic limit, so that we can choose to
symmetrize them (for instance in the fully symmetric irreducible
representation) in order to construct a uniform state. Equivalently,
one can choose one VBC state and then average over distances. To
perform these computations is a bit tedious but straightforward. We
give here some details on the calculation, and relevant results are
shown in Tables~\ref{table1} and \ref{table2}.

For spin correlations, if sites $i$ and $j$ belong to the same
hexagon, then correlations can be obtained from the hexagon
wavefunctions given in Eqs.~(\ref{2.02}), (\ref{2.04}), and (\ref{2.06}); if site $i$
corresponds to a polarized site (resp. resonating hexagon), then
$\langle S_i^z\rangle=1/2$ (resp. $(9m-3)/12$). If sites $i$ and $j$
are sufficiently distant, then correlation simply factorizes since we
have a product state. As an example for $m=7/9$, the nearest-neighbor
spin correlations is given by
$$ \left(\frac{1}{12}+2\times \frac{1}{3}\times\frac{1}{2}\right)/3- \left(\frac{m}{2}\right)^2 = -\frac{1}{81}$$
while at large distance we find only two different values
$$ \left( \frac{1}{3}\times\frac{1}{2}+\frac{1}{3}\times\frac{1}{2} +
\frac{1}{3}\times\frac{1}{3}\right) - \left(\frac{m}{2}\right)^2 =
-\frac{1}{324}, $$ and
$$ \left( \frac{1}{3}\times\frac{1}{3}+\frac{1}{3}\times\frac{1}{3} + \frac{1}{2}\times\frac{1}{2}\right) - \left(\frac{m}{2}\right)^2 = \frac{1}{162},$$
and similar computations can be performed for other $m$ and distances. 

Dimer correlation can be computed in a similar way, but since we have
many possibilities for $(ij)$ and $(k\ell)$ bonds, we will not give all
numbers. We need to compute $\langle (\boldsymbol{S}_i\cdot
\boldsymbol{S}_j)\rangle$: when $i$ and $j$ are nearest neighbors  
inside one resonating hexagon, then this is simply the energy per bond
for the wavefunctions (\ref{2.02}), (\ref{2.04}), and (\ref{2.06}); in
the other case, one site is necessarily a polarized one so that
correlations reduce to $\langle S_i^z S_j^z\rangle = \langle
S_i^z\rangle \langle S_j^z\rangle = (9m-3)/24$. Let us denote these
two values as $d_1$ and $d_2$.  If we neglect short-distance effects
that require detailed computation (ten different relative bond
positions), then computations are much simpler and we have found only
two cases
$$ (d_1 d_2 + 2d_2^2)/3 - \left(\frac{d_1+2d_2}{3}\right)^2$$
and 
$$ (d_1^2 + 2d_2^2)/3 - \left(\frac{d_1+2d_2}{3}\right)^2$$
As an example for $m=7/9$, we obtain respectively $-1/144$ and $1/72$. 
Some of these numbers are reported in Table~\ref{table2}. 

\smallskip

\renewcommand{\arraystretch}{1.2}
\begin{table}[tb!]
\begin{tabular}{|c||c|c|c||c|c|}
\hline
$\langle S_i^z S_j^z\rangle_c $ & \multicolumn{3}{|c||}{same hexagon} & \multicolumn{2}{|c|}{along one direction} \\ 
\hline
 & $d=1$ & $d=2$ & $d=3$ & $d=2$ & $d=3$\\
\hline
$m=7/9$ & $-1/81$ &  $-1/324$ & $-1/81$ & $-1/324$ & $1/162$\\
\hline
$m=5/9$ & $-0.045$ & $0.0167$ & $-0.005$ & $-1/81$ & $2/81$ \\
\hline
$m=1/3$ & $-0.0797$ & $0.0787$ & $-0.053$ & $-1/36$ & $1/18$\\
\hline
\end{tabular}
\caption{\label{table1}
Connected $\langle S_i^z S_j^z\rangle_c$ for $i,j$ inside one hexagon at Manhattan distance $d$ or along one direction for various VBC states corresponding to different $m$.}
\end{table}

\renewcommand{\arraystretch}{1.2}
\begin{table}[tb!]
\begin{tabular}{|c||c|c||c|c|}
\hline
$\langle (\boldsymbol{S}_i\cdot \boldsymbol{S}_j) (\boldsymbol{S}_k\cdot \boldsymbol{S}_\ell) \rangle_c $ & \multicolumn{2}{|c||}{same hexagon} & \multicolumn{2}{|c|}{different hexagon} \\ 
\hline
 & $d=1$ & $d=2$ & positive & negative\\
\hline
$m=7/9$ & $-5/216$ &  $-5/216$ & -1/144 & 1/72\\
\hline
$m=5/9$ &  0.06655 & 0.01630 & -0.04154 & 0.02192 \\
\hline
$m=1/3$ & 0.02974 & 0.07946 & -0.02424 & 0.04849\\
\hline
\end{tabular}
\caption{\label{table2}
Connected $\langle (\boldsymbol{S}_i\cdot \boldsymbol{S}_j) (\boldsymbol{S}_k\cdot \boldsymbol{S}_\ell)\rangle_c$ for a fixed reference bond $(i,j)$ and various bonds $(k,\ell)$ either belonging to the same hexagon at Manhattan distance $d$ or not. Other relative bond positions are possible so that twelve different values can be found. Here, we show only four different values for each VBC corresponding to different $m$.}
\end{table}

{\bf $\boldsymbol{m=7/9}$ plateau}: 
We observe degeneracies larger than three on small lattices presumably due to
the existence of short loops  around them, compare also
Ref.~\onlinecite{loc_mag_review2}.
However, using our largest $N=63a$ lattice which diameter is $d=\sqrt{21}$, we do confirm that the ground-state at $m=7/9$ is exactly three-fold degenerate (corresponding to the three possible VBC), so that one can form eigenstates with momentum $\Gamma$ and two-fold states with K. 
Our numerical correlations perfectly agree with our analytical results performed on the VBC states, and tiny differences can be attributed to the small overlap between the three magnon states on a finite lattice. Data are plotted in Fig.~\ref{fig:CorrPanel}, which is a perfect signature of the existence of a VBC state.

\smallskip

{\bf $\boldsymbol{m=5/9}$ plateau}: 

In this case, we consider lattice $N=36d$. Although spin and dimer correlations are a bit less intense than in the pure VBC state (we know that this is not an eigenstate anymore), both the sign patterns and the long-range order are in very good agreement. This leads us to the conclusion that for $m=5/9$ a VBC state emerges, and since this is a gapped state, we expect a finite plateau at this magnetization. 

\smallskip
{\bf $\boldsymbol{m=1/3}$ plateau}:

The same conclusion seems to be valid for $m=1/3$ where correlations have a similar pattern to the ones found in the pure VBC state. Both short-distance properties of the VBC wavefunction are recovered in ED data, but also the fact that correlations do not depend much on distance and seem to remain finite.  

\section{Conclusion}

We have shown that the kagome antiferromagnet in strong magnetic field
exhibits a non-trivial magnetization curve. Previous studies had
indicated that plateaux should exist for $m=1/3$ and $m=7/9$ of its
saturation value, but recently the existence of the 1/3 plateau has
been challenged.\cite{Nakano2010,Sakai2011a,Sakai2011b,Sakai2013}  
Here we have not only presented further support for the existence of the 1/3 plateau, but also evidence in favor of an $m=5/9$ plateau, in addition to the exactly known 7/9 plateau. We have presented a unified view of these plateaux states, which are valence bond crystals that break lattice symmetries.

Our approach is based on a generalization of the exact magnon-crystal
state which exists at $m=7/9$, that we believe captures as well the physics
for the other plateaux.
These wavefunctions correspond to simple VBC
states such as depicted in Fig.~\ref{fig01} where resonating hexagons
have a fixed magnetization equals to 0, $1$ and 2 respectively, and
they share similar properties.

Our Exact Diagonalizations on large lattices have confirmed (i) that
these three plateaux have a finite extent in the thermodynamic limit,
of widths roughly equal to $0.5\, J$, $0.3\, J$ and $0.1\, J$
respectively (ii) for these magnetizations, the ground-state is
three-fold degenerate and correspond to the expected VBC state. Last
but not least, both spin and dimer correlations are in very good
agreement with the magnon VBC state, which allows to present a simple
physical picture of these three gapped phases.

Since for $m=1/3$, it is possible to construct a featureless bosonic state on this lattice\cite{Parameswaran2013}
(i.e., a unique quantum Mott insulator of bosons that has no broken symmetry or topological order), it would
be interesting to look for models that can interpolate between having
a VBC ground-state or a featureless one. 
In this context, more work should be devoted to understand 
the difference between Heisenberg and bosonic models, 
and also to investigate whether supersolid phases are stable or not. 

Let us finally mention that  the localized-magnon scenario holds for a quite large variety of one-,
two- and three-dimensional frustrated
lattices.\cite{loc_mag_review1,loc_mag_review2} Hence, we may
argue that our VBC approach based on a generalization of the
exact magnon-crystal
state might be applicable to other lattices, such as the star lattice,
\cite{Richter2004,star04,star07,star_exp} the sorrel net,\cite{sorrel1,sorrel2} or the square-kagome
lattice.\cite{sorrel2,Sidd,squago}
Moreover, it is known  that also for  spin quantum                
numbers $S>1/2$ and for anisotropic XXZ antiferromagnets the
magnon-crystal state exists.\cite{Schulenburg2002,loc_mag_review1,loc_mag_review2}
Therefore, the investigation of  plateaux states for other lattices and/or
spin-$S$  XXZ models is a fruitful field for further studies.

\acknowledgments

Numerical simulations were performed at CALMIP and GENCI.
O.~D. would like to thank the Abdus Salam International Centre for 
Theoretical Physics (Trieste, Italy)
for partial support of these studies through the Senior Associate award.
AH and AML acknowledge support through  FOR1807 (DFG / FWF)
and JR and OD through RI 615/21-1 (DFG).

\section*{Note added}

After submission of this manuscript, we learned about   
Ref.~\onlinecite{Nishimoto2013} where magnetization plateaux are investigated with the
Density-Matrix Renormalization Group algorithm. These results agree with ours
for the three plateaux that we have investigated, namely that they correspond to valence bond crystals.
Moreover, Ref.~\onlinecite{Nishimoto2013} predicts an exotic quantum plateau at $m=1/9$.

\bigskip
\appendix

\section{Lattice geometries}

Since we are using several kinds of lattices, we give their definitions using ${\bf a}$ and ${\bf b}$ translations to define the
torus with periodic boundary conditions. Unit length corresponds to the Bravais
lattice unit, i.e., two lattice spacings. We define
the diameter of each lattice as $d=\min(|{\bf a}|, |{\bf b}|,|{\bf a}-{\bf b}|)$.  
In this work, we only use lattices with a K point in their Brillouin zone 
(shown in bold in Table \ref{tab:lattice}). In particular, in constrast to
Refs.~\onlinecite{Nakano2010,Sakai2011a,Sakai2011b,Sakai2013}, we do not
use 39-sites lattices since they are not compatible with the expected VBC order.

\begin{table}[t!]
\begin{tabular}{|c|c|c|c|}
\hline
Name & {\bf a} & {\bf b} & diameter \\
\hline
21 & $(5/2,\sqrt{3}/2)$ & $(1/2,3\sqrt{3}/2)$ & $\sqrt{7}$\\
24 & $(2,\sqrt{3})$ & $(-2,\sqrt{3})$ & $\sqrt{7}$\\
27a & $(5/2,\sqrt{3}/2) $ & $(-3/2,3\sqrt{3}/2)$ & $\sqrt{7}$\\
{\bf 27b} & $(3,0)$ & $(3/2,3\sqrt{3}/2)$& 3 \\
30 & $(7/2,\sqrt{3}/2)$ & $(1/2,3\sqrt{3}/2)$ & $\sqrt{7}$\\
33 & $(5/2,\sqrt{3}/2)$ & $(-1,2\sqrt{3})$ & $\sqrt{7}$\\
36a & $(4,0)$ & $(1/2,3\sqrt{3}/2)$ & $\sqrt{7}$\\
36b & $(4,0)$ & $(3/2,3\sqrt{3}/2)$ & 3 \\
{\bf 36c} & $(-3,\sqrt{3})$ & $(3/2,3\sqrt{3}/2)$ & 3 \\
{\bf 36d} & $(3,\sqrt{3})$ & $(0,2\sqrt{3})$ & $\sqrt{12}$\\
39a & $(5/2,\sqrt{3}/2$) & $(-3,2\sqrt{3})$ & $\sqrt{7}$ \\
39b & $(5/2,3\sqrt{3}/2)$  & $(-1,2\sqrt{3})$ & $\sqrt{13}$ \\
42a & $(1/2,3\sqrt{3}/2)$ & $(9/2,-\sqrt{3}/2)$ & $\sqrt{7}$ \\
42b & $(0,2\sqrt{3})$ & $(7/2,\sqrt{3}/2)$ & $\sqrt{12}$\\
45a & $(5,0)$ & $(3/2,3\sqrt{3}/2)$ & 3 \\
{\bf 45b} & $(3,2\sqrt{3})$ & $(-3/2,3\sqrt{3}/2)$ & 3\\
{\bf 54} & $(9/2,3\sqrt{3}/2)$ & $(-3/2,3\sqrt{3}/2)$ & 3\\
{\bf 63a} & $(9/2,\sqrt{3}/2)$ & $(3/2,5\sqrt{3}/2)$ & $\sqrt{21}$ \\
63b & $(7,0)$ & $(3/2,3\sqrt{3}/2)$ & 3 \\
\hline
\end{tabular}
\caption{\label{tab:lattice}Finite lattices studied in this work. Listed are the number of spins
$N$; the basis vectors {\bf a}, {\bf b} in the $xy$
plane; the diameter $d=\min(|{\bf a}|, |{\bf b}|,|{\bf a}-{\bf b}|)$.
Lattices shown in bold contain the K point in their Brillouin zone (see Fig.~\ref{fig:gaps} for definition). }
\end{table}


 
\end{document}